\documentstyle[aps
,preprint%
]{revtex}
\newcommand{\be}{\begin{equation}}
\newcommand{\ee}{\end{equation}}
\newcommand{\bea}{\begin{eqnarray}}
\newcommand{\eea}{\end{eqnarray}}
\newcommand{\nn}{\nonumber \\}
\newcommand{\Bx}{\mbox{\boldmath$x$}}

\newcommand{\Bv}{\mbox{\boldmath$v$}}
\newcommand{\BV}{\mbox{\boldmath$V$}}

\newcommand{\BL}{\mbox{\boldmath$L$}}
\newcommand{\Bnabla}{\mbox{\boldmath$\nabla$}}

\newcommand{\dd}{d}
\begin{document}
\draft
\preprint{AJC-HEP-28}
\date{\today
}
\title{
Extreme Dilatonic Black Holes on a Torus
}
\author{Kiyoshi~Shiraishi%
\thanks{e-mail: {\tt g00345@sinet.ad.jp,
shiraish@air.akita-u.ac.jp}
}}
\address{Akita Junior College\\
Shimokitade-sakura, Akita-shi, Akita 010, Japan
}
\maketitle
\begin{abstract}
The interaction of maximally charged dilatonic
black holes on $R^{4}\times T^{d}$ is studied
in the low-velocity limit.
In particular, the scattering of two black holes
on $R^{4}\times S^{1}$ is investigated.
\end{abstract}

\vspace{7mm}
\vfill
\eject

\section{Introduction}

In recent years, solitons in field theories have attracted
much attention. Solitonic objects may play important roles
in the unity of fundamental theories including string
theories~\cite{DKL}.
The relation between topology of the space and the solitonic objects
is a key subject for exploring the ultimate theory of everything
in higher dimensions.
The mechanism of the realization of the physical
three-dimensional space by compactification and the method of
detecting the
extra dimensions in our universe may be obtained by studying the
solitonic objects in higher dimensions.

The exact solution for an arbitrary number of
maximally charged dilatonic black holes has been found
by several authors~\cite{GHS,Shi1}.
Static multi-soliton solutions, which saturate certain
Bogomol'nyi-type bounds, have been studied
in various theories~\cite{DKL}.
There is no net static force among the solitonic objects described
by such a multi-soliton solution.
In the present paper, using the identification process,
we show the multi-centered solution on
the toroidally compactified spacetime, $R^{4}\times T^{d}$ in
Sec.~\ref{sec:sol}. The same method can be used
for general toroidal compactifications in arbitrary dimensions.
We find that the solution reduces to the multi-black hole
solution on  $R^{4}$ if the scale of the torus shrinks to zero.

Recently the present author has also studied the interaction of
maximally charged dilatonic black holes at low velocities
and obtained the moduli space metric
for two such black holes explicitly~\cite{Shi2,Shi3}.
In Sec.~\ref{sec:int} in the present paper,
we present the low-energy interaction of the multi-black hole system
on $R^{4}\times T^{d}$. We concentrate on the case with $d=1$:
This case arises in the five-dimensional
Einstein-Maxwell-dilaton
theory with Kaluza-Klein compactification on a circle.
Two-body scattering is analyzed in this case in Sec.~\ref{sec:two}.
The last section~\ref{sec:conc} is devoted to a brief conclusion.

\section{The exact multi-soliton solution on $R^{4}\times T^{\dd}$}
\label{sec:sol}

The action for the effective field theory of string theory in
$(d+4)$ dimensional spacetime, including $U(1)$ gauge field but
discarding the antisymmetric field strength $H_{\mu\nu\lambda}$, is
\be
S=\int d^{d+4}x\ \frac{\sqrt{-g}}{16\pi G}\ e^{-2\phi}\left[ R +
4 (\nabla \phi)^{2} - F^{2} \right] +
\mbox{(surface terms)}\; ,
\ee
where $R$ is the scalar curvature and $\phi$ stands for the dilaton.
The vector field $A_{\mu}$ is related to
the field strength $F_{\mu\nu}$ by
$F_{\mu\nu}=\partial_{\mu}A_{\nu}-\partial_{\nu}A_{\mu}$,
where $\mu$, $\nu$ run over $0,1,\ldots,d+3$.
$G$ is the Newton's constant.

Rescaling the action as
$g_{\mu\nu}\rightarrow e^{\frac{4}{d+2}\phi} g_{\mu\nu}$
yields an action for the Einstein-Maxwell-dilaton theory,
which includes the Einstein-Hilbert action for gravitation:
\be
S=\int d^{d+4}x\ \frac{\sqrt{-g}}{16\pi G} \left[ R -
\frac{4}{d+2} (\nabla \phi)^{2} - e^{-\frac{4}{d+2}\phi} F^{2}
\right] +
\mbox{(surface terms)}\; .
\ee
We take the action of this form to study the solution and
interaction of the solitonic objects derived from it in this paper.

The static multi-centered solution with flat asymptotic region
$R^{d+4}$ takes the form~\cite{GHS,Shi1}
\be
ds^{2}= - V^{-2\frac{d+1}{d+2}}dt^{2} + V^{\frac{2}{d+2}} d\Bx^{2},\;
A_{\mu}dx^{\mu} = \frac{1}{\sqrt{2}}\left(1-\frac{1}{V}\right)dt,\;
e^{-\frac{4}{d+2}\phi}=V^{\frac{2}{d+2}}\; ,
\label{eq:met}
\ee
where
\be
V=1+\sum_{\alpha} \frac{4\pi}{A_{d+2}}
\frac{2 G m_{\alpha}}{(d+1)|\Bx -\Bx_{\alpha}|^{d+1}}\; .
\label{eq:sol}
\ee
where $A_{d+2}=\frac{2\pi^{(d+3)/2}}{\Gamma ((d+3)/2)}$.
The solution describes the situation that the $\alpha$-th
nonrotating, charged dilatonic black hole
in the extreme limit with mass $m_{\alpha}$
is located at $\Bx =\Bx_{\alpha}$.%
\footnote{Strictly speaking, there are
naked singularities in the solution.
Nevertheless, we use the term ``black hole'' because the extreme case
may still have generic properties of black holes in terms of
their classical dynamics.}


Now we consider the multi-black hole solution on a torus space.
Although we can consider arbitrary dimensions, we concentrate
on the case of $R^{4}\times T^{d}$, with a physical interest.
To derive the solution which globally has
the topology of $R^{4}\times T^{d}$ from the asymptotically
flat solution
(\ref{eq:met},\ref{eq:sol}), we use the technique of identification.
We denote the coordinate of $d$-dimensional subspace as
$\xi_{i}$ ($i=1,2,\ldots d$).
We obtain the torus space by identifying
$\xi_{i}$ and $\xi_{i}+L_{i}\ell_{i}$, where $\ell_{i}$ is an integer.
$L_{i}$ is the circumference of the $i$-th direction of the torus.
As for the multi-centered solution, the copious images of
a black hole at $\Bx_{\alpha}$ must be located at $\Bx_{\alpha}+\BL$
with $\BL\equiv
(L_{1}\ell_{1}, L_{2}\ell_{2}, \ldots , L_{d}\ell_{d}, 0, 0, 0)$
and have the same mass $m_{\alpha}$.

Consequently, on the space with toroidal compactification,
the set of solutions for the metric and the other fields
takes the same form as Eq.~(\ref{eq:met}), but $V$ has the form
\be
V=1+\sum_{\alpha}
\frac{2 \tilde{G} m_{\alpha}}{r_{\alpha}}
\sum_{\ell_{1}=-\infty}^{\infty}\cdots
\sum_{\ell_{d}=-\infty}^{\infty}
\exp\left(-2\pi r_{\alpha}\sqrt{\sum_{i=1}^{d}\frac{\ell_{i}^{2}}
{L_{i}^{2}}}\right)
\prod_{i=1}^{d}\cos \ell_{i}\theta_{\alpha,i}\; ,
\label{eq:solonT}
\ee
where $r_{\alpha}=
\sqrt{(x-x_{\alpha})^{2}+(y-y_{\alpha})^{2}+(z-z_{\alpha})^{2}}$
and $\theta_{\alpha,i}=2\pi (\xi_{i}-\xi_{\alpha,i})/L_{i}$.
$\tilde{G}$ represents an effective Newton constant defined by
$\tilde{G}=G/{\cal V}_{T}$ with ${\cal V}_{T}=\prod_{i=1}^{d}L_{i}$.

In the limit of all $L_{i}\rightarrow 0$, $V$ is reduced to
\be
V=1+\sum_{\alpha}
\frac{2 \tilde{G} m_{\alpha}}{r_{\alpha}}\; ,
\label{eq:sol4}
\ee
which is the same as the solution
for the four dimensional case~\cite{GHS,Shi1}.

For $d=1$, the sum over $\ell$ in Eq.~(\ref{eq:solonT})
can be done analytically,
and then $V$ is obtained as follows:
\be
V=1+\sum_{\alpha}
\frac{2 \tilde{G} m_{\alpha}}{r_{\alpha}}
\frac{\sinh 2\pi r_{\alpha}/L}{\cosh 2\pi r_{\alpha}/L -
\cos \theta_{\alpha}}\; .
\label{eq:sol5}
\ee

The multi-soliton solution on a torus
in Einstein-Maxwell-dilaton theory
with an arbitrary dilaton coupling~\cite{Shi1} can be constructed
in the same manner.

\section{Interactions}
\label{sec:int}

The present author has applied the method of
Ferrell and Eardley~\cite{FerEar} to calculate
the interaction energy of
the maximally charged dilatonic black holes
at low velocities by making use of the exact,
static solution~(\ref{eq:met}).
The equations in detail for its derivation
can be referred to the previous work~\cite{Shi2,Shi3}.

Since there are only two-body (velocity-dependent) forces
in the multi-black hole system in this case,%
\footnote{For a general dilaton coupling, the interaction
has more complicate form~\cite{Shi2,Shi3}.}
the general expression for the $O(v^2)$ lagrangian of the system
of an arbitrary configuration of black holes
can be easily obtained as:~\cite{Shi2,Shi3}
\bea
L_{v^2}&=&\sum_{\alpha}\frac{1}{2}m_{\alpha}\Bv_{\alpha}^{2} +
\sum_{\alpha\beta}\frac{4\pi}{A_{d+2}}
\frac{G m_{\alpha}m_{\beta} |\Bv_{\alpha}-\Bv_{\beta}|^{2}}
{2(d+1)|\Bx_{\alpha}-\Bx_{\beta}|^{d+1}}\nn
&=&\frac{1}{2}M\BV^{2} +
\sum_{\alpha\beta}
\frac{m_{\alpha}m_{\beta} |\Bv_{\alpha}-\Bv_{\beta}|^{2}}{4 M}
\left(1+\frac{4\pi}{A_{d+2}}
\frac{2 G M}{(d+1)|\Bx_{\alpha}-\Bx_{\beta}|^{d+1}}\right)\;,
\label{eq:int}
\eea
where $\Bv_{\alpha}$ is the velocity of
the extreme dilatonic black hole with mass $m_{\alpha}$.
The total mass $M$ is given by $M=\sum_{\alpha} m_{\alpha}$,
and $\BV$ is the velocity of
the center of mass; $\BV\equiv
{\sum_{\alpha} m_{\alpha} \Bv_{\alpha}}/{M}$.
We have disregarded the constant term in the lagrangian,
which is proportional to the total mass of the system.


Again we use the identification of the images to construct
the interaction lagrangian of extreme dilatonic black holes
on $R^{4}\times T^{d}$.
The images of a certain black hole have the same velocity as well as
the same mass.
We also note that the sum of the interaction lagrangian of
the images yields an overall infinite factor,
$\sum_{\ell_{1}=-\infty}^{\infty}\cdots
\sum_{\ell_{d}=-\infty}^{\infty}1$.
We have simply to omit this overall factor.

Finally we get the following lagrangian of $O(v^{2})$:
\bea
& &\tilde{L}_{v^2}=\frac{1}{2}M\BV^{2}\nn
& &+\sum_{\alpha\beta}
\frac{m_{\alpha}m_{\beta} |\Bv_{\alpha}-\Bv_{\beta}|^{2}}{4 M}
\left[1+\frac{2 \tilde{G} M}{\rho_{\alpha\beta}}
\sum_{\ell_{1}=-\infty}^{\infty}\cdots
\sum_{\ell_{d}=-\infty}^{\infty}
\exp\left(-2\pi\rho_{\alpha\beta}\sqrt{\sum_{i=1}^{d}
\frac{\ell_{i}^{2}}{L_{i}^{2}}}\right)
\prod_{i=1}^{d}\cos \ell_{i}\theta_{\alpha\beta,i}\right]\; ,\nn
{}
\label{eq:intonT}
\eea
where $\rho_{\alpha\beta}=
\sqrt{(x_{\alpha}-x_{\beta})^{2}+(y_{\alpha}-y_{\beta})^{2}+
(z_{\alpha}-z_{\beta})^{2}}$ and
$\theta_{\alpha\beta}=2\pi (\xi_{\alpha,i}-\xi_{\beta,i})/L_{i}$.%
\footnote{For a case of a general dilaton coupling,
since the interaction
contains many-body, velocity-dependent forces~\cite{Shi2,Shi3},
the expression will be more complicated. The resemblance between
Eq.~(\ref{eq:intonT}) and Eq.~(\ref{eq:solonT}) is a coincidence.}
Note that we have {\em not} used any long-distance approximation.
The dependence of the configuration is exact at this order in $v^2$.

In the next section, we study the two-body scattering
on $R^{4}\times S^{1}$ using the low-energy interaction lagrangian.

\section{Two-body scattering on $R^{4}\times S^{1}$}
\label{sec:two}

The slow motion of solitons can be described as
geodesics on the space of parameters (moduli space)
for static solutions.
Therefore the calculation of the metric on moduli space
is the most effective way to
investigate the interaction of slowly moving solitons
in classical field theory~\cite{Manton}.
In this section, we consider the scattering of two maximally charged
dilatonic black holes
on $R^{4}\times S^{1}$ using the metric of moduli space.

The $O(v^{2})$ lagrangian of the two-body system
of the black holes with masses $m_{A}$ and $m_{B}$ is obtained from
Eq.~(\ref{eq:intonT}), that is
\be
\tilde{L}_{v^2}=\frac{1}{2}(m_{A}+m_{B})\BV^{2}+
\frac{1}{2}\mu
\left[1+\frac{2 \tilde{G} (m_{A}+m_{B})}{\rho}
\frac{\sinh 2\pi\rho/L}{\cosh 2\pi\rho/L -
\cos \theta}\right]|\Bv_{A}-\Bv_{B}|^{2}\;,
\label{eq:int5}
\ee
where  $\rho=\rho_{AB}$ and $\theta=\theta_{AB}$.
The reduced mass $\mu$ is given by
$\mu=m_{A}m_{B}/(m_{A}+m_{B})$.

Hereafter we assume that the black holes move
in a (scattering) plane in the three-dimensional space.
Thus the moduli space of this configuration
is reduced to be a three-dimensional space
parameterized by the mutual distance $\rho$,
the azimuthal angle $\varphi$,
and the angle difference on the extra circle $\theta$.

For this two-body system, the metric on moduli space of
relative motion can be read from Eq.~(\ref{eq:int5}) as
\be
ds_{MS}^2=\gamma(\rho,\theta)
\left(d\rho^{2}+\rho^{2}d\varphi^{2}+\frac{L^{2}}{4\pi^{2}}
d\theta^{2}\right)\; ,
\label{eq:le1}
\ee
with
\be
\gamma(\rho,\theta)=1+\frac{2 \tilde{G} M}{\rho}
\frac{\sinh 2\pi\rho/L}{\cosh 2\pi\rho/L -\cos \theta}\; ,
\label{eq:le2}
\ee
where $M=m_{A}+m_{B}$.


We describe the two-body scattering using the coordinate
where one of the extreme dilatonic black holes is located
at the origin.
We set a length scale such that $2\tilde{G}M=1$ in the followings.
The parameters used in the description of the scattering process
is exhibited in FIG.~\ref{fig1}.
The effect of the extra dimensions becomes important when
the typical, or minimum distance between the black holes $\rho_0$
and the compactification scale $L$ are the same order.
Therefore, we choose $L=\pi$ in the following numerical
calculations to see the effect well.


First we examine the case that $\theta$ is fixed
to be zero everywhere.
This situation satisfies the equation of motion for $\theta$
derived from the geodesic equation on the moduli space.
The orbit of the projectile is determined by the geodesic equation
derived from the line element~(\ref{eq:le1},~\ref{eq:le2})
by the variational method.
The relation between the impact parameter $b$ and
the scattering angle $\Theta$ is shown in FIG.~\ref{fig2} for
$L=\pi$.
Note that the horizontal axis indicates $1/b$.
If $b$ is sufficiently large, the deflection is described
by the Rutherford scattering~\cite{Shi3}.
For large $b$, the angle of deflection is well described by the
expression for four-dimensional black holes~\cite{Shi3,Shi2}:
\be
\Theta=2\arctan\frac{1}{2b}\; .
\label{eq:lb}
\ee
If $b$ is small and comparable to $L$,
since the shape of the moduli space in the vicinity of the origin
looks like the one of the five-dimensional black hole system, then
the scattering angle $\Theta$ diverges when $b$ approaches
a certain value $b_{crit}$~\cite{Shi2}.
The two dilatonic black holes coalescence for
$b<b_{crit}$~\cite{Shi2}.
The critical value $b_{crit}$ is given by
\be
b_{crit}=L/\pi \ ,
\ee
in the case on $R^{4}\times S^{1}$.
For small $b$, the scattering angle is well approximated by the
expression for five-dimensional black holes~\cite{Shi2}:
\be
\Theta=\frac{1-\sqrt{1-\frac{L}{b^{2}\pi}}}%
{\sqrt{1-\frac{L}{b^{2}\pi}}}\pi\; .
\label{eq:sb}
\ee


Next we consider the effect of the relative motion of black holes
on the extra space $S^{1}$.
We take $\rho_{0}=1$,
and at this position we assume non-zero value of
the gradient of $\theta$ along the path,
that is $|\Bnabla\theta|$. For simplicity,
we set the initial value of $\theta$ ($=\theta_{init}$)
at this point equals to zero.
We show an example of the trajectory of
an extreme dilatonic black hole
when $L=\pi$, $\rho_{0}=1$, and the initial value of
$|\Bnabla\theta|=|\Bnabla\theta|_{init}=4$ in FIG.~\ref{fig3}.
As one can see from the behavior in FIG.~\ref{fig3},
the value of the angle $\varphi_0$ turns out to be very
sensitive to the initial value of $|\Bnabla\theta|$,
though the effect is suppressed if $\rho_{0}$
is much larger than $L$.
This is due to the periodic nature of the interaction on $\theta$
and thus the motion is {\em not} merely a projection of the path
of two-body scattering in the isotropic five dimensions.
The periodicity of the effective force may bring about chaotic motion
of black holes.
More deep inspection will be needed and studied elsewhere.

\section{Conclusion}
\label{sec:conc}

We have shown the multi-soliton solution
in Einstein-Maxwell-dilaton theory on $R^{4}\times T^{d}$, and
investigated the interaction of maximally charged dilatonic
black holes on $R^{4}\times T^{d}$ in the low velocity limit.
The scattering of two black holes on $R^{4}\times S^{1}$
has been studied numerically.

These features found in the case of $R^{4}\times S^{1}$
may exist also in a general dimensional case.
The effect of the extra dimensions can be found
in the process of the ``near-head-on'' collision.
The effect of the relative motion in the extra space is also
observable in our three-dimensional world
if the typical scale of the two-body system is the same order of
the compactification scale.

We must bear in mind that our analyses are based on the
low-velocity lagrangian. The terms of higher order in $v$
and the radiational back-reaction may become important if all the
typical scales of the scattering process are very small.
These corrections will be studied in the future work.

The true effective theory of string theory contains
the antisymmetric tensor field $B_{\mu\nu}$ as well as
the higher-order terms of the curvature and the field strength.
The analysis on the theory including these will be
of much importance if we pursue the connection between
string theory and solitons~\cite{DKL,DR,KM,CMP}.



\newpage

\begin{figure}
\caption{A schematic view of the two-body scattering process
and an introduction of parameters.}
\label{fig1}
\end{figure}


\begin{figure}
\caption{The scattering angle $\Theta$ versus
the inverse of the impact parameter $b$, for $L=\pi$.
Two curves described by Eq.~(13) and Eq.~(15)
are also exhibited as dashed lines.}
\label{fig2}
\end{figure}


\begin{figure}
\caption{An example of the trajectory of
an extreme dilatonic black hole
when $L=\pi$, $\rho_{0}=1$, and the initial value of
$|\nabla\theta|_{init}=4$, and $\theta_{init}=0$.}
\label{fig3}
\end{figure}

\end{document}